\documentclass[twocolumn,prd,aps,nofootinbib,superscriptaddress]{revtex4-2}
\usepackage{amsmath, amssymb}
\usepackage{graphicx}
\usepackage{hyperref}
\usepackage{booktabs}
\usepackage{tikz}
\usepackage{pgfplots}
\pgfplotsset{compat=1.18}
\usepgfplotslibrary{fillbetween}

\begin{document}

\title{LIGO, LISA and Ultralight Axion-like Dark Matter}

\author{Lawrence M. Krauss}
\affiliation{The Origins Project Foundation, Phoenix AZ 85027}
\email{krauss@asu.edu}

\vspace*{1cm}
\date{\today}

\begin{abstract}
Interferometric searches for axion-like dark matter (ALP DM) via the induced photon birefringence in a coherent background have been proposed and analyzed by several authors over the past decade, with tabletop experiments (LIDA, ADBC, DANCE) reporting first results and detailed calculations available for both ground-based (LIGO, GEO600 via DarkGEO) and space-based (LISA, TianQin, Taiji, BBO via Gu\'e et al. and Yao et al.) gravitational-wave detectors.  We revisit this class of measurements within the framework of Krauss (2019), and provide a detailed treatment of the interplay between the DM coherence length $\lambda_{\rm coh}$, the interferometer arm length $L$, and the detector transfer function, with attention to the distinct suppression regimes $\lambda_{\rm coh} > L$ (cancellation) and $\lambda_{\rm coh} < L$ (random walk).  For LIGO the sensitivity is set by the Fabry-P\'erot arm cavity response peaking sharply near the cavity free spectral range, reproducing the results of Nagano \textit{et al.}; the naive single-pass estimate substantially overstates the reach and cannot be applied directly.  For LISA, with the polarization-sensitive modification proposed by Gu\'e \textit{et al.} or Yao \textit{et al.}, the natural mass range $m_a \sim 4\times10^{-19}$--$4\times10^{-16}$~eV (LISA science band) is accessible and the projected reach $g_{a\gamma\gamma}\lesssim {\rm few}\times 10^{-13}$~GeV$^{-1}$ (realistic optical-metrology noise) improves on the CAST bound over most of the band by up to about two orders of magnitude.  We discuss directional signatures from the Galactic DM wind, both annual and sidereal, as a discrimination tool for candidate signals.  Local DM substructure with anisotropic velocity dispersion can shift and enhance the reach in specific mass windows.
\end{abstract}

\maketitle

\section{Introduction}

Axions, resulting from the proposed Peccei-Quinn solution of the strong CP problem~\cite{pq}, could, due to spontaneous symmetry breaking in the early universe, arise as a coherent classical dark matter background field today~\cite{dm}.  While QCD axions are strongly constrained by a combination of terrestrial and astrophysical bounds, the possibility of other similar pseudo-goldstone boson cosmic backgrounds that may comprise the dark matter in galaxies remains of interest.

Anomaly considerations suggest such pseudoscalar fields $a$ are likely to have an axion-like photon coupling of the form
\begin{equation}
  \mathcal{L} \approx -\tfrac{1}{4}\,g_{a\gamma\gamma}\, a \,F_{\mu\nu}\tilde{F}^{\mu\nu}.
\end{equation}
For conventional QCD axions there is a fixed relationship between $g_{a\gamma\gamma}$ and the axion mass $m_a$, but for a generalized pseudoscalar background this need not be the case.

A coherent axion-photon coupling gives rise to a rich phenomenology.  The most extensively developed detection channels are microwave-cavity conversion in a static magnetic field~\cite{sikivielmk,DM-Radio,ABRACADABRA,MADMAX}.  Beyond these, the axion background induces a birefringence of the vacuum: circularly polarized photons of opposite helicity acquire a small time-varying effective mass squared of opposite sign,
\begin{equation}
  (\delta m_\gamma^{\pm})^2(t) = \mp\,2\,g_{a\gamma\gamma}\,\omega\,\dot a(t),
  \label{eq:effmass}
\end{equation}
which follows from the modified Amp\`ere equation $\nabla\times\vec B = \dot{\vec E} + g_{a\gamma\gamma}\dot a\,\vec B$ in the spatially-uniform limit~\cite{Krauss-2019,renau}.  Equivalently, linearly polarized light undergoes a time-varying polarization rotation.  This birefringence effect has been exploited in pulsar polarization array measurements~\cite{LiuSmoot,Parkes-PPA-2024,EPTA-PPA-2025} and in astrophysical polarimetry~\cite{Ivanov,Fujita-Tazaki}.

The possibility of using laser interferometry to search for this birefringence effect --- and thereby probe ALP dark matter --- has an extensive history.  While interferometric probes of this effect were suggested in passing in~\cite{Krauss-2019}, DeRocco and Hook~\cite{DeRocco-Hook} had previously proposed using interferometry of circularly polarized light with quarter-wave plates inserted at each mirror to preserve the polarization on reflection, thereby avoiding the cancellation of the signal on normal-incidence reflection discussed in Sec.~\ref{sec:phaseandcavity}; this scheme was shown to potentially reach a few orders of magnitude beyond then-current constraints.  A parallel line of development uses birefringent optical cavities, either in ring or bowtie geometry to circumvent the mirror-reflection problem naturally, developed theoretically by Obata, Fujita and Michimura~\cite{Obata-2018}, Liu \textit{et al.}~\cite{Liu-Elwood} and Martynov and Miao~\cite{Martynov-Miao}, and realized experimentally by DANCE~\cite{DANCE}, ADBC~\cite{ADBC} and LIDA~\cite{LIDA}, all of which have reported first data.  Adaptation of existing GW infrastructure has been proposed by Nagano \textit{et al.}~\cite{Nagano-2019,Nagano-2021} for Advanced LIGO and Cosmic Explorer, and by the DarkGEO collaboration~\cite{DarkGEO} for GEO600.  Extension to space-based interferometers has been analyzed by Gu\'e, Hees and Wolf~\cite{Gue-2025} for LISA-like detectors (with a single modification: replacement of one half-wave plate by a circular polarizer), and by Yao, Jiang and Tang~\cite{Yao-2025} using Sagnac time-delay-interferometry combinations.  A heterodyne readout with photon-counting has been analyzed by Ref.~\cite{heterodyne-photon}. Constraints on scalar dark matter from LIGO have also been reported~\cite{LIGO-scalarDM}.

The present manuscript revisits this line of investigation within the framework of~\cite{Krauss-2019}, and also addresses three additional aspects that have not received a unified treatment in the earlier literature: (i) the interplay between the DM coherence length $\lambda_{\rm coh}$ and the interferometer arm length $L$, distinguishing the ``cancellation'' regime $\lambda_{\rm coh}>L$ from the ``random-walk'' regime $\lambda_{\rm coh}<L$; (ii) how the Fabry-P\'erot arm-cavity response of LIGO-like detectors~\cite{Nagano-2019,DeRocco-Hook}, which determines the accessible mass range, combines with the coherence-length suppression; and (iii) the annual and sidereal wind-modulation discrimination signatures in the interferometric birefringence channel, together with the effect of local anisotropic substructure.  For LISA-like detectors, we compare the projected reach quoted in~\cite{Gue-2025,Yao-2025} against the shot-noise-limited estimates in the~\cite{Krauss-2019} framework.  Throughout, we project sensitivity limits as curves in the $(m_a, g_{a\gamma\gamma})$ plane rather than along the QCD axion line.

\section{Photon phase shift and cavity response}
\label{sec:phaseandcavity}

With $a(t) = a_0\cos(m_a t)$ and $a_0 = \sqrt{2\rho_{\rm DM}}/m_a$ fixed by the local DM energy density, and for $(\delta m_\gamma^{\pm})^2 \ll \omega^2$, a probe photon of either circular polarization accumulates a phase along a single-pass baseline $L$
\begin{equation}
  \delta\Phi_\pm(t) \approx \pm\,g_{a\gamma\gamma}\,(m_a a_0)\,L\,\sin(m_a t)
   \;\equiv\; \pm\beta\sin(m_a t),
\end{equation}
with
\begin{equation}
  \beta \;\equiv\; g_{a\gamma\gamma}\,\sqrt{2\rho_{\rm DM}}\,L,
  \label{eq:beta}
\end{equation}
independent of the probe-photon frequency $\omega$.  For a spatially-uniform background, the signal appears at a sideband $\Omega = \pm m_a$ relative to the laser carrier with amplitude $\beta/2 \ll 1$.  A small finite momentum, due to velocity dispersion in our galaxy $\delta v$, spreads the sideband over a fractional bandwidth $\Delta\Omega/\Omega \sim \delta v^2 \sim 10^{-6}$.

\paragraph{Mirror-reflection cancellation in Fabry-P\'erot cavities.}
The single-pass expression above is not directly applicable to a Fabry-P\'erot arm cavity such as those in Advanced LIGO.  In such a cavity, the photon reflects off end mirrors at normal incidence.  Reflection reverses the propagation direction $\hat k \to -\hat k$ but leaves the physical spin (E-field rotation direction) unchanged, so the helicity $h = \hat S \cdot \hat k$ flips sign.  Since the axion-induced effective mass has a helicity-dependent sign, the accumulated phase reverses on the return trip and nearly cancels that of the forward trip.  What survives is a residual of order $m_a L/c$, so that the round-trip phase scales as $\beta \times (m_a L/c)$ rather than the naive $\beta \times \text{finesse}$~\cite{DeRocco-Hook,Nagano-2019}.  (Note that this cancellation formally requires DM-field coherence between the forward and return trips, i.e.\ $\tau_{\rm coh} > 2L/c$.  For all existing and proposed detectors this is satisfied by at least several orders of magnitude, and this effect would only be relevant for hypothetical detectors of much longer arm length or higher target masses.)  The full cavity transfer function has a broadband suppression by $(m_a L/c)^2 \sim (f_a/f_{\rm FSR})^2$ at frequencies well below the free spectral range $f_{\rm FSR} = c/(2L)$, and exhibits a sharp resonant peak at $f_a = f_{\rm FSR}$ where the cavity response function has its maximum~\cite{Nagano-2019}.

Two classes of solutions circumvent this cancellation.  In DeRocco and Hook~\cite{DeRocco-Hook}, quarter-wave plates are inserted at the end mirrors so that the polarization state is flipped in a way that undoes the helicity reversal.  In a high-power detector such as Advanced LIGO, however, transmissive optics inside the arm cavities would add substantial intra-cavity loss and a large thermal lens that would be very difficult to compensate; this practical obstacle is a principal reason why all existing tabletop experiments use ring or bowtie geometries instead.  Alternatively, bowtie or ring cavities of the sort employed by DANCE~\cite{DANCE}, ADBC~\cite{ADBC}, and LIDA~\cite{LIDA} circulate the light without normal-incidence reflection.  For Advanced LIGO, Nagano \textit{et al.}~\cite{Nagano-2019} derive the full arm-cavity transfer function and show that, without a modification, the potential sensitivity peaks sharply at $g_{a\gamma\gamma}\simeq 10^{-14}$~GeV$^{-1}$ at the FSR resonance $m_a \simeq 1.56\times 10^{-10}$~eV, and drops to a broadband floor near $g_{a\gamma\gamma}\simeq 8\times 10^{-13}$~GeV$^{-1}$ over a wider mass range around $m_a \simeq 3\times 10^{-13}$~eV.  A follow-up~\cite{Nagano-2021} using the arm-cavity transmitted beam relaxes some of these constraints.

\paragraph{Single-pass configurations: LISA and DarkGEO-II/III.}
For space-based interferometers such as LISA, TianQin, Taiji, and BBO, the interspacecraft laser link is single-pass: there is no reflection off a distant mirror, and the photon travels one way from the sending spacecraft to the receiving spacecraft.  Consequently the mirror-reflection cancellation is absent, and the naive single-pass phase shift~\eqref{eq:beta} applies to each link.  However, as shown by Yao \textit{et al.}~\cite{Yao-2025}, the standard LISA design uses linearly polarized light and its current phase-metrology chain is insensitive to axion-induced birefringence at $\mathcal{O}(g_{a\gamma\gamma})$: the axion signal appears only as an $\mathcal{O}(g_{a\gamma\gamma}^2)$ amplitude modulation of the horizontally-polarized component.  Sensitivity to the axion field at leading order therefore requires a modification of the optical bench.  Gu\'e, Hees, and Wolf~\cite{Gue-2025} propose replacing one half-wave plate by a circular polarizer, a minimal change.  Yao \textit{et al.}~\cite{Yao-2025} propose a more involved arrangement using quarter- and half-wave plates and analyze the response in the Sagnac time-delay-interferometry combinations.  In what follows we assume such a modification is in place and analyze the resulting sensitivity.  For ground-based ring cavities, DarkGEO-II and later stages~\cite{DarkGEO} avoid the cancellation entirely.

\paragraph{Two-arm geometry.}
Existing gravitational-wave detectors, and planned next-generation terrestrial detectors, are differential Michelson interferometers.  This topology is essential for gravitational waves, which change the two arm lengths differentially, but it offers little for an axion-induced birefringence signal, which affects both arms nearly equally: the dominant noise sources in a detector such as LIGO (quantum, coating-thermal and seismic noise) are largely uncorrelated between the arms, so common-mode rejection brings no significant advantage.  Our reason for analyzing the differential response in Sec.~\ref{sec:differential} is different.  It is the response a gravitational-wave detector delivers without a change of topology, and Sec.~\ref{sec:differential} quantifies the price paid for it: for an axion field that is nearly uniform across the apparatus, the differential subtraction removes the DM signal to zeroth order in $L/\lambda_{\rm coh}$, leaving a residual that scales as $L/\lambda_{\rm coh}$ from the gradient of the DM field.  An order-unity differential response is recovered only when $\lambda_{\rm coh}\lesssim L$, or through the geometric asymmetry of the two arms with respect to the DM wind discussed in Sec.~\ref{sec:directional}.  A single-arm polarimetric readout, or a ring cavity, avoids this suppression altogether.  For LISA, the Sagnac TDI combination adopted by~\cite{Yao-2025} effectively converts the two-way differential structure into an equivalent ring geometry.

\paragraph{Summary of configurations.}
Table~\ref{tab:configs} summarizes the detector configurations discussed in this manuscript.  We consider these four because together they span the options for adapting existing or planned gravitational-wave infrastructure (a, b, d), together with the purpose-built cavity geometry (c) that currently sets direct interferometric limits.  In any multi-arm configuration [(a), (b), (d)], a differential readout between arms provides sensitivity to the direction of the DM wind (Sec.~\ref{sec:directional}), at the cost of the suppression discussed in Sec.~\ref{sec:differential}.

\begin{table}[t]
\caption{Interferometric configurations for detecting axion-induced birefringence considered in this manuscript.}
\label{tab:configs}
\footnotesize
\setlength{\tabcolsep}{3pt}
\begin{tabular}{p{0.30\columnwidth}p{0.30\columnwidth}p{0.31\columnwidth}}
\toprule
Configuration & Merits & Drawbacks \\
\midrule
(a) Linear Fabry-P\'erot arm with intra-cavity waveplates~\cite{DeRocco-Hook}
  & Broadband response in an existing km-scale cavity
  & Transmissive optics in high-power arms: large loss and thermal lensing \\[3pt]
(b) Linear Fabry-P\'erot arm, polarization readout, no waveplates~\cite{Nagano-2019,Nagano-2021}
  & Minimal modification; compatible with GW operation
  & Mirror-reflection cancellation; sharp sensitivity only near $f_{\rm FSR}$ \\[3pt]
(c) Ring or bowtie cavity~\cite{LIDA,ADBC,DANCE,DarkGEO}
  & No normal-incidence reflection; broadband; operating now
  & Short (m-scale) arms, or a new km-scale facility; oblique-incidence phase shifts must be compensated \\[3pt]
(d) Single-pass space link with polarization optics~\cite{Gue-2025,Yao-2025}
  & No reflection; $10^{9}$~m arms; small optical-bench change
  & Limited by optical-metrology noise; requires a change to the LISA design \\
\bottomrule
\end{tabular}
\end{table}

\section{Two observable schemes}

The birefringence effect~\eqref{eq:effmass} admits two complementary readout schemes in a Michelson-type interferometer, both discussed extensively in the earlier literature.  We summarize them here for completeness.

\paragraph{(i) Circular-polarization phase readout.}  Inject circularly
polarized light into the interferometer.  Each arm imparts a phase
$\delta\Phi_\pm = \pm\beta\sin(m_a t)$ to the chosen helicity; the standard
interferometric output is sensitive to the differential phase between arms.
For ground-based Fabry-P\'erot cavities, quarter-wave plates must be inserted at each mirror to preserve the circular polarization on reflection, following DeRocco and Hook~\cite{DeRocco-Hook}; as noted in Sec.~\ref{sec:phaseandcavity}, this is a challenging modification in a high-power arm cavity.  For LISA and other space-based interferometers, the single-pass link geometry means that only a modification of the optical bench (a circular polarizer replacing a half-wave plate~\cite{Gue-2025}, or an appropriately arranged waveplate stack~\cite{Yao-2025}) is required.

\paragraph{(ii) Polarimetric interferometer.}  Keep linear polarization
through the optics.  After the recombination beamsplitter, place a linear
polarization analyzer at $45^\circ$ to the input polarization.  The
transmitted intensity then encodes the differential polarization rotation
$\Delta\theta = (\beta_1 - \beta_2)\sin(m_a t)$ \emph{linearly} in
$g_{a\gamma\gamma}$.  This scheme is essentially that of Nagano \textit{et al.}~\cite{Nagano-2019} and is the interferometric analogue of a Pulsar Polarization Array~\cite{LiuSmoot,Parkes-PPA-2024, EPTA-PPA-2025}.

At leading order in $g_{a\gamma\gamma}$ both schemes give the same
sensitivity to the differential phase, experience the same spatial
coherence-length scaling (see below),
and saturate the same shot-noise floor.  The choice between them is
practical; we carry the analysis through in terms of the differential
phase $\Delta\Phi$ between arms without committing to either scheme.

\section{Differential interferometer response}
\label{sec:differential}

As explained in Sec.~\ref{sec:phaseandcavity}, we analyze the differential response because it is what the Michelson topology of existing gravitational-wave detectors delivers; this section quantifies the suppression that results.  In both of these schemes the role of the dark-matter coherence length
$\lambda_{\rm coh} = 2\pi/(m_a\delta v)$ in setting the differential
interferometer response deserves careful treatment.

A Michelson interferometer is sensitive only to differences between its two
arms, exhibiting common-mode rejection of any signal that affects both arms
equally.  (It is worth noting, however, that in LIGO the dominant noise sources are largely uncorrelated between the arms, so common-mode rejection offers little benefit for this signal.  The advantage of the differential measurement lies elsewhere: by cancelling the leading, direction-independent part of the signal, it isolates the contribution from the gradient of the DM field, and hence is sensitive to the direction of the dark-matter wind, as discussed in Sec.~\ref{sec:directional}.  This makes it most useful for characterizing a signal once a detection has been made.)  The differential observable is
\begin{equation}
  \Delta\Phi(t) \;=\; \delta\Phi_1(t) - \delta\Phi_2(t).
\end{equation}

When the dark-matter coherence length exceeds the arm length,
$\lambda_{\rm coh} \gg L$, the DM field is approximately uniform across the
apparatus to zeroth order in $L/\lambda_{\rm coh}$ and both arms experience
the same time-varying mass.  The leading non-vanishing differential
contribution comes from the DM gradient between the arm endpoints, which
scales as $L/\lambda_{\rm coh}$ (Figure~\ref{fig:suppression}).  Modelling the local DM as a plane wave with
wavevector $k_{\rm DM} = m_a\,\delta v$ and averaging over the ensemble of DM-field realizations (equivalently, over directions and over many coherence times), one finds
\begin{equation}
  \frac{\Delta\Phi_{\rm rms}}{\beta} \;\sim\; \frac{L}{\lambda_{\rm coh}}
  \qquad (\lambda_{\rm coh} > L).
  \label{eq:cancellation}
\end{equation}

In the opposite limit $\lambda_{\rm coh} \ll L$, each arm samples
$N = L/\lambda_{\rm coh}$ independent coherence patches with random phases.
The differential phase is a random walk of $\mathcal{O}(\sqrt{N})$ steps of
amplitude $g_{a\gamma\gamma}\sqrt{2\rho_{\rm DM}}\lambda_{\rm coh}$:
\begin{equation}
  \frac{\Delta\Phi_{\rm rms}}{\beta}
   \;\sim\; \sqrt{\frac{\lambda_{\rm coh}}{L}}
   \qquad (\lambda_{\rm coh} < L).
  \label{eq:walk}
\end{equation}

The two branches~\eqref{eq:cancellation} and~\eqref{eq:walk} meet at
$\lambda_{\rm coh} = L$, where the differential response peaks at unity.

\begin{figure}[t]
\centering
\begin{tikzpicture}
\begin{axis}[
    width=8.5cm, height=6cm,
    xmode=log, ymode=log,
    xlabel={$\lambda_{\rm coh}/L$},
    ylabel={$\Delta\Phi_{\rm rms}/\beta$},
    xmin=1e-3, xmax=1e3,
    ymin=1e-3, ymax=2,
    grid=both, minor grid style={gray!10}, major grid style={gray!25},
    tick label style={font=\small}, label style={font=\small},
]
\addplot[blue, thick, samples=80, domain=1e-3:1]{sqrt(x)};
\addplot[blue, thick, samples=80, domain=1:1e3]{1/x};
\addplot[red, dashed, samples=2, domain=1:1] coordinates {(1,1e-3) (1,2)};
\node at (axis cs:0.05,0.05) [font=\footnotesize, anchor=south west]
   {random walk};
\node at (axis cs:20,0.05) [font=\footnotesize, anchor=south west]
   {cancellation};
\node at (axis cs:1,1.5) [font=\footnotesize, color=red, anchor=south]
   {peak};
\end{axis}
\end{tikzpicture}
\caption{Suppression of the differential interferometric response as a
function of the dark-matter coherence length relative to the arm length.
The signal peaks at $\lambda_{\rm coh}=L$ with two distinct decay regimes:
random-walk suppression $\sqrt{\lambda_{\rm coh}/L}$ when the coherence
length is shorter than the arms, and common-mode cancellation
$L/\lambda_{\rm coh}$ when it is longer.}
\label{fig:suppression}
\end{figure}

\section{Signal-to-noise and the linewidth-limited reach}

For a probe-photon flux $\dot N$ and an observation time $T \gg T_{\rm coh}
= 1/(m_a \delta v^2)$, the shot-noise-limited signal-to-noise in the
sideband at $\Omega = m_a$ is
\begin{equation}
  \mathrm{SNR}^2 \;\propto\;
    \Delta\Phi_{\rm rms}^2\,\dot N\,\sqrt{T\,T_{\rm coh}}.
\end{equation}

Two regimes of temporal coherence are relevant.  When $T_{\rm obs}<T_{\rm coh}$
the search is in the fully-coherent regime: the signal is essentially
monochromatic over the observation, matched filtering gives
$\mathrm{SNR}^2\propto T_{\rm obs}$, and the sensitivity improves linearly
with integration time.  When $T_{\rm obs}>T_{\rm coh}$ the search becomes
linewidth-limited: the signal has fractional bandwidth $\Delta f/f\sim 1/Q
\sim \delta v^2$ from the dark-matter velocity dispersion, and coherent
integration saturates after $T_{\rm coh}$.  Additional observation time
enters only through incoherent stacking of the power spectrum across
$N = T_{\rm obs}/T_{\rm coh}$ independent coherence-time segments, giving
the $\sqrt{T_{\rm obs}\cdot T_{\rm coh}}$ scaling above.  Crucially, the broadening
of the signal in frequency space (the $1/T_{\rm coh}$ linewidth) and the
spatial coherence-length suppression $\Delta\Phi_{\rm rms}/\beta$ both
descend from the same underlying velocity dispersion $\delta v$, and both
must be retained simultaneously when projecting the reach.

Substituting~\eqref{eq:cancellation} in the cancellation regime and solving
for the minimum detectable coupling at $\mathrm{SNR}=1$,
\begin{equation}
  \boxed{\;
    g_{a\gamma\gamma,\min} \;\simeq\;
      \frac{\lambda_{\rm coh}(m_a, \delta v)}{L}\cdot
        \frac{1}{\sqrt{2\rho_{\rm DM}}\,L}\cdot
        \frac{1}{\sqrt{\dot N\,\sqrt{T\,T_{\rm coh}}}}.
    \;}
  \label{eq:gmin}
\end{equation}

In the cancellation regime ($\lambda_{\rm coh} \gg L$),
$\lambda_{\rm coh}/L \propto 1/(m_a\delta v)$ and $\sqrt{T_{\rm coh}}\propto
1/\delta v$, so that $g_{a\gamma\gamma,\min} \propto \delta v^{-1/2}$ at fixed
$m_a$.  

Details of an axion halo are not certain and a variety of theoretical analyses have been carried out
(i.e. \cite{clusters,structure}). In the cancellation regime, colder streams enhance $T_{\rm coh}$ but
correspondingly worsen the cancellation suppression; the cancellation loss dominates and search sensitivity degrades.  In the random-walk regime (heterodyne section below), the opposite is true: cold streams shift the peak mass upward and \emph{improve} the optimum reach.  Caustics, in which $\delta v \to 0$, suppress the
differential signal to zero by common-mode cancellation in either case.

\section{LISA sensitivity in its native science band}

Throughout this section we assume that the polarization-sensitive modification of the LISA optical bench described in Sec.~\ref{sec:phaseandcavity}~\cite{Gue-2025,Yao-2025} has been implemented; without it, the standard LISA readout is insensitive to the birefringence signal at first order in $g_{a\gamma\gamma}$.

For LISA's baseline $L = 2.5\times 10^9$~m and signal frequencies
$f_{\rm sig} = m_a/(2\pi\hbar)$ within the science band $0.1$~mHz to $0.1$~Hz,
the corresponding relevant ALP mass range is
\begin{equation}
  4\times 10^{-19}~\text{eV} \lesssim m_a \lesssim 4\times 10^{-16}~\text{eV}.
\end{equation}
In Standard Halo Model dark matter ($\delta v\sim 10^{-3}c$), the coherence
length across this mass range is $\lambda_{\rm coh}\sim 3\times 10^{12}$ to
$3\times 10^{15}$~m, well above the LISA arm length, placing the search
squarely in the cancellation regime.

For a shot-noise limited detection with unsqueezed light, laser power $P = 1$~W at $\omega = 1$~eV (giving a photon flux $\dot N \approx 6\times 10^{18}$~s$^{-1}$) and an observation time of $T = 1$~yr, equation~\eqref{eq:gmin} predicts a shot-noise-limited reach

\begin{center}
\begin{tabular}{cccc}
\toprule
$m_a$ [eV] & $f_{\rm sig}$ [Hz] &
$L/\lambda_{\rm coh}$ & $g_{a\gamma\gamma,\min}^{\rm shot}$ [GeV$^{-1}$] \\
\midrule
$4\times 10^{-19}$ & $10^{-4}$ & $8\times 10^{-7}$ & $\sim 3.3\times 10^{-12}$ \\
$4\times 10^{-18}$ & $10^{-3}$ & $8\times 10^{-6}$ & $\sim 3.3\times 10^{-13}$ \\
$4\times 10^{-17}$ & $10^{-2}$ & $8\times 10^{-5}$ & $\sim 3.9\times 10^{-14}$ \\
$8\times 10^{-17}$ & $2\times 10^{-2}$ & $1.6\times 10^{-4}$
   & $\sim 2.3\times 10^{-14}$ \\
$4\times 10^{-16}$ & $10^{-1}$ & $8\times 10^{-4}$ & $\sim 6.9\times 10^{-15}$ \\
\bottomrule
\end{tabular}
\end{center}

These shot-noise-limited numbers set a fundamental lower bound but do not represent the realistic sensitivity, which is set by LISA's optical-metrology-system (OMS) noise floor $s_{\rm oms} \simeq 15$~pm/$\sqrt{\rm Hz}$ and by acceleration noise from test-mass interactions.  Using the realistic LISA noise model of Babak, Hewitson, and Petiteau~\cite{Babak-noise}, Yao \textit{et al.}~\cite{Yao-2025} (using Sagnac TDI combinations) and Gu\'e \textit{et al.}~\cite{Gue-2025} (using a single circular-polarizer modification) obtain a projected sensitivity
\begin{equation}
  g_{a\gamma\gamma}^{\rm LISA} \simeq 3\times 10^{-13}~\text{GeV}^{-1} 
  \quad \text{at} \quad m_a \sim 10^{-17}~\text{eV},
\end{equation}
which improves upon the CAST bound by $\sim 200\times$ but is weaker than the shot-noise-limited estimates in our table above by up to about two orders of magnitude, depending on mass.  The physical reason is that OMS noise dominates over shot noise across the LISA science band.  We nonetheless list the shot-noise-limited numbers to establish the fundamental scale.

The CAST helioscope bound $g_{a\gamma\gamma} \lesssim 6.6\times 10^{-11}~$GeV$^{-1}$~\cite{CAST} and tighter astrophysical bounds --- MWD polarisation~\cite{MWDPolarisation} at $g_{a\gamma\gamma}\lesssim 5.5\times 10^{-12}$~GeV$^{-1}$, Chandra observations of NGC 1275~\cite{Chandra-NGC1275} at $g_{a\gamma\gamma}\lesssim 6.3\times 10^{-13}$~GeV$^{-1}$ for $10^{-14}<m_a<10^{-11}$~eV, and black-hole superradiance exclusion strips (model-dependent) --- occupy the parameter space above the LISA reach across most of the mass range.  See Fig.~\ref{fig:sensitivity} for the full landscape.

Beyond the reach quoted above, next-generation space-based interferometers with tighter noise budgets, such as Big Bang Observer, could push the reach several orders of magnitude further~\cite{Yao-2025}.  The BBO projection is $g_{a\gamma\gamma} \sim 10^{-16}$~GeV$^{-1}$ across a broad mass range.

\section{LIGO with the arm-cavity transfer function}
\label{sec:LIGO}

Advanced LIGO uses Fabry-P\'erot arm cavities of length $L=4$~km with finesse $\mathcal{F}\sim 450$, giving a free spectral range $f_{\rm FSR} = c/(2L)\simeq 37.5$~kHz.  As discussed in Sec.~\ref{sec:phaseandcavity}, the cavity response function to an axion signal is not that of a naive single-pass baseline: mirror-reflection cancellation suppresses the axion-induced polarization phase away from the cavity resonance, with a broadband suppression by $\sim (f_a/f_{\rm FSR})^2$ at $f_a \ll f_{\rm FSR}$ and a resonant peak at $f_a = f_{\rm FSR}$~\cite{DeRocco-Hook,Nagano-2019}.

Applying the full cavity treatment, Nagano \textit{et al.}~\cite{Nagano-2019} find that an Advanced LIGO-like detector (with an appropriate polarization readout implemented as a modification of the standard configuration) has a potential sensitivity to the axion-photon coupling
\begin{align}
  g_{a\gamma\gamma}^{\rm LIGO} &\simeq 8\times 10^{-13}~\text{GeV}^{-1}
    \quad (m_a \sim 3\times 10^{-13}~\text{eV}), \nonumber\\
  g_{a\gamma\gamma}^{\rm LIGO} &\simeq 10^{-14}~\text{GeV}^{-1}
    \quad (m_a = 1.56\times 10^{-10}~\text{eV}).
\end{align}
The first (broadband) reach is roughly two orders below the CAST helioscope bound; the second, at the FSR resonance, reaches a further two orders lower.  These projections assume a 1-year observation and include only the shot noise of an unsqueezed vacuum~\cite{Nagano-2019}.  Squeezed light of the kind already in use in Advanced LIGO would improve them: since $g_{a\gamma\gamma,\min}$ scales as the noise amplitude, $x$~dB of effective squeezing improves the reach by a factor $10^{x/20}$, i.e.\ about a factor of 2 for 6~dB.  The naive single-pass estimate based only on $\beta = g_{a\gamma\gamma}\sqrt{2\rho_{\rm DM}}L$, which ignores the mirror-reflection cancellation, cannot be applied to the arm cavity: it overstates the cavity response away from the FSR and misses the resonant structure altogether.

As a representative next-generation facility we adopt the Cosmic-Explorer-like configuration of Ref.~\cite{Nagano-2019}, with 40~km arms (corresponding to its FSR resonance at $m_a = 1.56\times 10^{-11}$~eV) and the same unsqueezed shot-noise assumption.  It pushes the broadband reach to $g_{a\gamma\gamma}\sim 4\times 10^{-14}$~GeV$^{-1}$ around $m_a \sim 2\times 10^{-15}$~eV and its FSR peak to $\sim 8\times 10^{-17}$~GeV$^{-1}$ at $m_a = 1.56\times 10^{-11}$~eV.  Other planned next-generation detectors, such as the Einstein Telescope, could be treated in the same way; their reach would depend on their arm length, and hence FSR, and on their arm-cavity configuration.

The tabletop experiments LIDA~\cite{LIDA}, ADBC~\cite{ADBC}, and DANCE~\cite{DANCE} use bowtie or rectangular ring geometries that avoid normal-incidence reflection entirely, and hence do not suffer the FSR-limited response of a linear cavity; they trade a much smaller arm length ($L \sim 0.5$--5~m) for a broadband response.  First observing runs have set direct constraints in the mass ranges near $m_a \sim 2\times 10^{-9}$~eV~\cite{LIDA} and $10^{-12}$~eV~\cite{DANCE}.

We conclude that a LIGO-based search for the birefringence signal is best framed as a resonant peak near the FSR (with the Nagano \textit{et al.}~\cite{Nagano-2019} numbers) rather than as a broadband audio-frequency probe.  A heterodyne readout implemented at the transmitted-beam port~\cite{Nagano-2021} may access additional mass ranges but requires careful treatment of the cavity response and is beyond the scope of the present work.

\section{Heterodyne extension: LISA}

A shot-noise-limited estimate suggests that an RF heterodyne readout implemented on the transmitted-beam ports of the LISA spacecraft could in principle extend the effective mass range accessible to the birefringence signal well into the $\mu$eV regime.  Assuming a $40$~GHz photodetector bandwidth (matching commercially available telecom-grade photodetectors and optical phase-lock loops), and a shot-noise-limited detection at each frequency bin, the corresponding formal reach in the random-walk regime $\lambda_{\rm coh}<L$ is

\begin{center}
\begin{tabular}{cccc}
\toprule
$m_a$ [eV] & $f_{\rm sig}$ [Hz] & regime & $g_{a\gamma\gamma,\min}^{\rm shot}$ [GeV$^{-1}$] \\
\midrule
$5\times 10^{-13}$ & $1.2\times 10^2$ & peak & $3.3\times 10^{-17}$ \\
$10^{-12}$         & $2.4\times 10^2$ & random walk & $5.5\times 10^{-17}$ \\
$10^{-11}$         & $2.4\times 10^3$ & random walk & $3.2\times 10^{-16}$ \\
$10^{-10}$         & $2.4\times 10^4$ & random walk & $1.8\times 10^{-15}$ \\
$10^{-8}$          & $2.4\times 10^6$ & random walk & $5.5\times 10^{-14}$ \\
$10^{-6}$          & $2.4\times 10^8$ & random walk & $1.8\times 10^{-12}$ \\
$10^{-4}$          & $2.4\times 10^{10}$ & RW (bw limit) & $5.5\times 10^{-11}$ \\
\bottomrule
\end{tabular}
\end{center}

\emph{Note that these are pedagogical shot-noise-limited estimates.}  Several considerations qualify the numbers.  First, as in the LISA-native band, the realistic sensitivity is set by the noise floor of the readout chain and may be orders of magnitude weaker than the shot-noise floor.  Second, a heterodyne-photon-counting scheme has been analyzed by Ref.~\cite{heterodyne-photon} with a careful treatment of the RF chain; the achievable reach is set by the excess noise of the photon counting apparatus.  Third, the assumption of a fixed LO setting integrated for the full observation time $T$ is appropriate for a focused search at a specific target $m_a$.  A blind survey across the full heterodyne mass range with the same total observation time must divide that time among $N_{\rm bins} \sim \Delta\omega_{\rm LO,max}/B_{\rm RO}$ scan steps; the per-bin sensitivity then degrades as $g_{a\gamma\gamma,\min}^{\rm scan} = g_{a\gamma\gamma,\min}\cdot N_{\rm bins}^{1/4}$.  For $B_{\rm RO}\sim 1$~MHz and $\Delta\omega_{\rm LO,max}/(2\pi) = 40$~GHz, the survey penalty is $N_{\rm bins}^{1/4} \sim 15$.

\begin{figure*}[t]
\centering
\includegraphics[width=\textwidth]{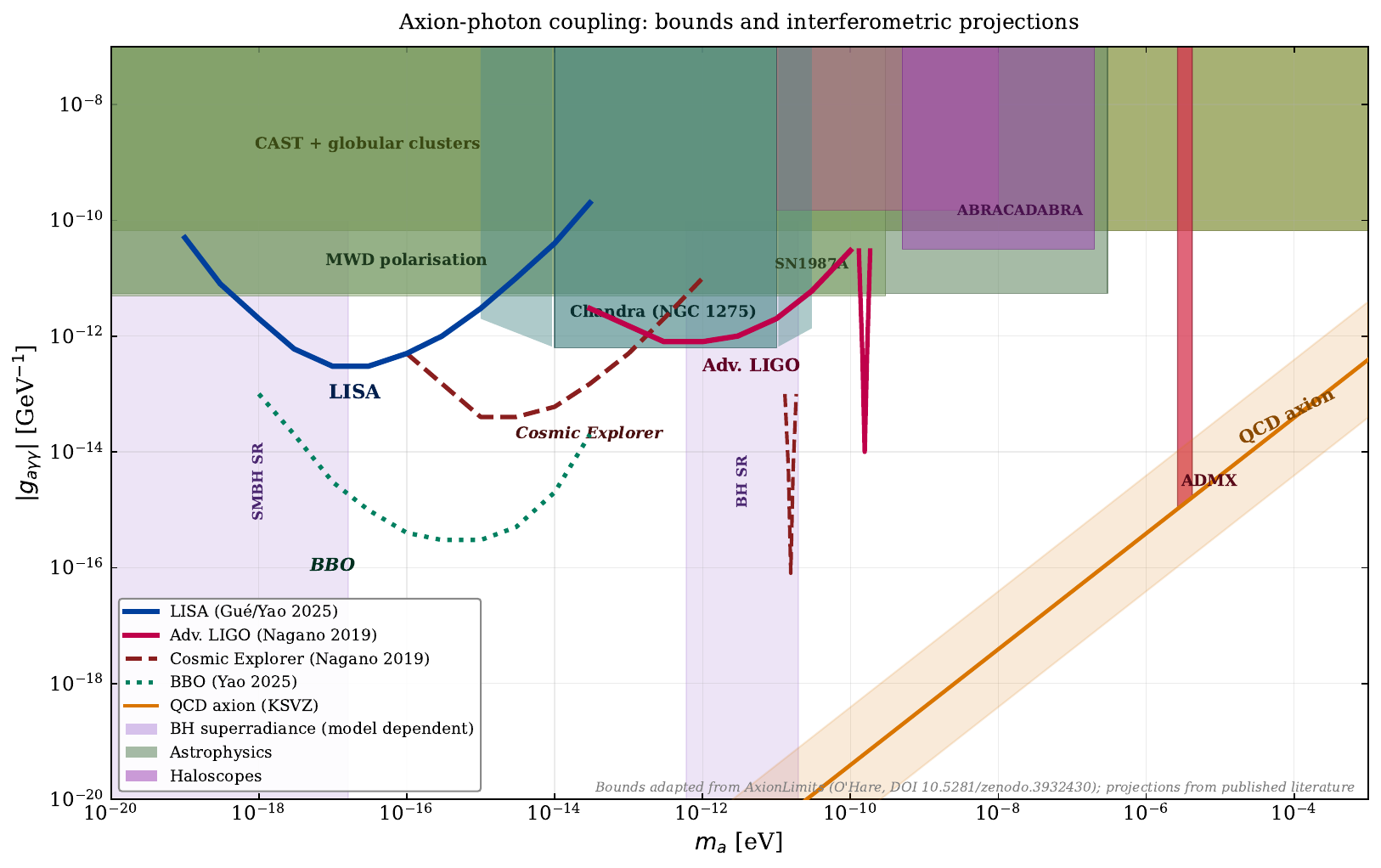}
\caption{Realistic projected sensitivity of interferometric axion-photon coupling searches, overlaid on the existing exclusion landscape.  Curves: LISA (with polarization-sensitive optical-bench modification per~\cite{Gue-2025,Yao-2025}); Advanced LIGO (with polarization readout per~\cite{Nagano-2019}, showing broadband floor and the sharp Fabry-P\'erot FSR resonance at $m_a = 1.56\times 10^{-10}$~eV); the 40-km Cosmic-Explorer-like configuration of Nagano et al.~\cite{Nagano-2019}, with correspondingly lower FSR peak at $m_a = 1.56\times 10^{-11}$~eV; and Big Bang Observer projection (Yao et al.~\cite{Yao-2025}).  The LIGO and Cosmic Explorer curves include only the shot noise of an unsqueezed vacuum; the LISA and BBO curves are not shot-noise estimates but use the full optical-metrology and acceleration noise models of Refs.~\cite{Yao-2025,Babak-noise}.  Shaded excluded regions: CAST helioscope and horizontal-branch stellar cooling ($g_{a\gamma\gamma}\lesssim 6.6\times 10^{-11}$~GeV$^{-1}$); magnetic white-dwarf polarisation~\cite{MWDPolarisation} ($\lesssim 5.5\times 10^{-12}$~GeV$^{-1}$); Chandra observations of NGC~1275~\cite{Chandra-NGC1275} ($\lesssim 6.3\times 10^{-13}$~GeV$^{-1}$ for $10^{-14}<m_a<10^{-11}$~eV); SN1987A; ABRACADABRA-10cm; ADMX; and black-hole superradiance exclusion strips from supermassive and stellar-mass BH spin measurements (vertical bands, model-dependent).  The QCD axion model line and broader QCD band are shown for reference.  Bounds collated from AxionLimits~\cite{AxionLimits}; projections from published literature.  LISA and Advanced LIGO both probe new parameter space in specific mass windows; Cosmic Explorer and BBO push several orders below current bounds in future.}
\label{fig:sensitivity}
\end{figure*}

We include the shot-noise-limited table above to establish the fundamental scale of the sensitivity achievable in principle, and to identify the mass ranges where an RF-heterodyne LISA-based measurement could complement the astrophysical bounds shown in Fig.~\ref{fig:sensitivity}.  A more realistic projection would require detailed noise modelling of the augmented LISA optical bench.  We suggest that this be considered as one of several possible extensions worthy of further study.

\section{Directional signatures from the Galactic dark-matter wind}
\label{sec:directional}

The interferometric birefringence signal considered in this manuscript carries characteristic annual and sidereal modulation signatures arising from the motion of the Earth through the local Galactic dark-matter wind.  To our knowledge, these signatures have not been developed in detail in the earlier interferometric axion-photon literature~\cite{DeRocco-Hook,Nagano-2019,Nagano-2021,Gue-2025,Yao-2025}.  We present the relevant expressions here as a discrimination tool for candidate signals.

The local dark-matter halo carries a bulk flow with respect to the Sun
arising from the Sun's motion through the Galactic rest frame (toward
Cygnus, $|\vec v_\odot| \approx 220$~km/s $\sim 7.3\times 10^{-4}c$).  In
the Standard Halo Model (SHM) with isotropic Maxwellian dispersion
centered on this flow, the velocity distribution in the Earth's rest
frame is
\begin{equation}
  f(\vec v) \propto \exp\!\left[-\frac{(\vec v -
    \vec v_{\rm wind}(t))^2}{v_0^2}\right],
\end{equation}
with $v_0 \approx |\vec v_\odot|$ and a time-dependent wind direction
$\vec v_{\rm wind}(t) = \vec v_\odot + \vec v_\oplus(t)$ that varies with
both Earth's orbital phase and rotational phase.  This produces several
calculable signatures whose presence can help confirm
any candidate detection.

\paragraph{Annual modulation.}  Earth's orbital velocity ($v_\oplus
\approx 30$~km/s) adds vectorially to the solar motion through the halo
with relative angle $\gamma \approx 60^\circ$ between Earth's orbital
plane and the Galactic flow direction, giving
\begin{equation}
  |\vec v_{\rm wind}|^2 = v_\odot^2 + v_\oplus^2
   + 2v_\odot v_\oplus \cos\!\big[2\pi(t-t_0)/\text{yr}\big]\cos\gamma,
\end{equation}
with peak in early June (Earth's velocity component aligned with the
Galactic flow) and trough in early December.  The peak-to-trough amplitude
is $\Delta v/v \approx 2 v_\oplus \cos\gamma/v_\odot \approx 14\%$.
The DM-induced sideband acquires a corresponding annual frequency shift
from the Doppler-shifted kinetic-energy term in the DM dispersion relation:
\begin{equation}
  \frac{\Delta f_{\rm sig}}{f_{\rm sig}} \;\sim\;
    \frac{v_\oplus v_\odot \cos\gamma}{c^2} \;\sim\; 10^{-7}.
\end{equation}
This shift is small compared to the source linewidth $1/Q \sim 10^{-6}$
but has the characteristic 1-year period and known phase, decoupled from
any terrestrial systematic.

\paragraph{Sidereal modulation.}  For a ground-based detector, Earth's
rotation modulates the projection of $\hat v_{\rm wind}$ onto the local
detector arms with sidereal period 23~h 56~m.  The two perpendicular arms
of a Michelson interferometer cannot both align with $\hat v_{\rm wind}$
simultaneously: when one is aligned, the other is orthogonal.  This
geometric asymmetry contributes a non-zero differential signal between the
arms even when both are individually in the cancellation regime
$\lambda_{\rm coh}>L$, with a sidereal-period amplitude modulation of
order unity (factor of $\sim 2$ peak-to-trough, depending on detector
latitude and arm orientation).  This orientation dependence is what gives the differential (Michelson) readout analyzed in Sec.~\ref{sec:differential} a physical handle on the DM signal even when the field is coherent across the apparatus.  The pattern is fully determined by the
detector's geographic location and the known Galactic-flow direction.

\paragraph{Cross-detector correlation.}  Detectors at different latitudes
and longitudes (LIGO Hanford, Livingston, and Virgo on Earth; LISA in
heliocentric orbit) sweep different orientations relative to
$\hat v_{\rm wind}$ at different times.  A genuine DM signal can show the
predicted sidereal-modulation pattern at each detector, with relative
phases set by the detector locations.  Random instrumental noise has no
such inter-detector correlation.  Cross-correlation of detector outputs
at the DM sideband frequency, weighted by the wind-direction template,
sharply discriminates DM from terrestrial systematics.

\paragraph{Anisotropic substructure.}  The conclusion changes if a
substantial fraction of local DM resides in cold streams or substructure
with intrinsic anisotropic velocity dispersion $\delta v_\parallel \ll
v_{\rm wind}$.  In that case the coherence length along the stream
direction can be substantially enhanced relative to the SHM value, and an
arm aligned with the stream direction could see effective coherence over
its full length while the transverse arm cannot.  Real streams in the
Milky Way (Sagittarius, GD-1, S2) carry $\lesssim 1\%$ of the local
density and have $\delta v_\parallel \lesssim 10^{-5}c$; their
contribution to a phase-modulation search would be subdominant in
amplitude but could produce a sharp narrowband signal with strong
directional dependence, providing both a discrimination opportunity and
a stream-velocity diagnostic if a signal is found.

\section{Discussion}

Our analysis reproduces, in the framework of~\cite{Krauss-2019}, several results that appear in the earlier interferometric-axion literature.  For LISA, a shot-noise-limited estimate yields peak reach $g_{a\gamma\gamma}\sim 7\times 10^{-15}$~GeV$^{-1}$ at $m_a\sim 4\times 10^{-16}$~eV; realistic noise budgets~\cite{Gue-2025,Yao-2025} give a more conservative $\sim 3\times 10^{-13}$~GeV$^{-1}$ at $m_a\sim 10^{-17}$~eV.  The realistic projection improves on the CAST helioscope bound by up to about two orders of magnitude across most of the LISA science band (the shot-noise estimate by up to four); both fall short of the QCD axion line.  Any measurement of the LISA-native sensitivity to the birefringence signal will require the polarization-sensitive modification of the optical bench proposed by Gu\'e \textit{et al.}~\cite{Gue-2025} or Yao \textit{et al.}~\cite{Yao-2025}.

For LIGO, the Fabry-P\'erot arm-cavity response function dominates the effective sensitivity.  The Nagano \textit{et al.}~\cite{Nagano-2019} analysis shows that Advanced LIGO can achieve a broadband reach of $g_{a\gamma\gamma}\sim 8\times 10^{-13}$~GeV$^{-1}$ around $m_a \sim 3\times 10^{-13}$~eV, with a sharp resonant peak reaching $\sim 10^{-14}$~GeV$^{-1}$ at the FSR frequency $m_a = 1.56\times 10^{-10}$~eV.  Cosmic Explorer-like next-generation ground-based detectors extend both by roughly an order of magnitude~\cite{Nagano-2019}.  The bowtie ring cavities employed by LIDA, ADBC, and DANCE avoid the FSR-limited response entirely and provide broadband sensitivity at their respective (much smaller) arm lengths~\cite{LIDA,ADBC,DANCE}.  The naive single-pass estimate is not a substitute for the full cavity treatment.  A single-arm polarimetric readout, which avoids the common-mode cancellation at the cost of reduced common-mode noise rejection, is a natural direction for detector-specific follow-up.

The LISA-native search has several attractive features once the polarization-sensitive modification is implemented.  First, the observable is a signal already present in LISA's primary measurement chain, and the search reduces to matched-filter peak finding in the science-band Fourier spectrum across the appropriate $Q\sim 10^6$ linewidth.  Second, the signal frequency provides a direct measurement of the mass with no dependence on the detector geometry beyond the overall $L/\lambda_{\rm coh}$ suppression.  Third, the annual and sidereal modulation signatures discussed above provide clean discrimination tools against terrestrial systematics.  A signal detected in LISA's band with consistency across multiple observation epochs, amplitude consistent with virial DM at the local density, and the predicted annual/sidereal modulation pattern would be strongly suggestive of a true ALP DM origin.

Several caveats merit mention.  LISA's noise floor at frequencies above the nominal science band is not yet fully characterised; realistic noise modelling gives a reach weaker than the shot-noise estimate by up to about two orders of magnitude, depending on mass.  The proposed heterodyne-readout extension of the birefringence measurement to higher masses is likewise a shot-noise-limited estimate, not a realistic projection; more careful readout-chain modelling is required.

Note also that the LISA limits overlap in part with claimed black-hole superradiance exclusion strips shown in Fig.~\ref{fig:sensitivity}.  These limits are model dependent and have known systematic astrophysical uncertainties; a direct interferometric probe of this region would therefore be useful.

\section{Conclusion}

A coherent background of axion-like dark matter (ALP) modifies conventional electrodynamics in several ways that provide opportunities to detect such a background.  Interferometric searches for the induced photon birefringence have been developed by many authors over the past decade~\cite{DeRocco-Hook,Obata-2018,Liu-Elwood,Nagano-2019,Martynov-Miao,Nagano-2021,DarkGEO,Gue-2025,Yao-2025,heterodyne-photon}, and tabletop implementations (LIDA, ADBC, DANCE~\cite{LIDA,ADBC,DANCE}) have reported first results.  The present manuscript revisits this class of measurements in the framework of~\cite{Krauss-2019}, focusing on the interplay between the DM coherence length and the interferometer arm length, the essential role of Fabry-P\'erot cavity response for LIGO-like detectors, and the discrimination signatures arising from the Galactic DM wind.

For LISA, with the polarization-sensitive optical-bench modification proposed by Gu\'e \textit{et al.}~\cite{Gue-2025} or Yao \textit{et al.}~\cite{Yao-2025}, we confirm a projected reach of $g_{a\gamma\gamma}\sim 3\times 10^{-13}$~GeV$^{-1}$ near $m_a \sim 10^{-17}$~eV, improving on the CAST helioscope bound by up to about two orders of magnitude while sitting above the tighter astrophysical bounds (MWD polarisation, Chandra NGC 1275) in parts of the range.  The shot-noise-limited estimates in the~\cite{Krauss-2019} framework are formally more optimistic by up to about two orders of magnitude, depending on mass, and represent the fundamental lower bound on the achievable sensitivity for an idealized readout chain.

For LIGO, the Fabry-P\'erot arm-cavity response function suppresses the axion birefringence signal by $(f_a/f_{\rm FSR})^2$ at frequencies well below the free spectral range $f_{\rm FSR} \simeq 37.5$~kHz.  The Nagano \textit{et al.}~\cite{Nagano-2019} treatment gives a broadband reach of $g_{a\gamma\gamma}\sim 8\times 10^{-13}$~GeV$^{-1}$ near $m_a \sim 3\times 10^{-13}$~eV and a sharp resonant peak at $\sim 10^{-14}$~GeV$^{-1}$ at the FSR frequency.  A naive single-pass estimate substantially overstates the achievable sensitivity.

A heterodyne-photodetector upgrade to either LIGO or LISA could in principle extend the reach in mass into the sub-$\mu$eV and $\mu$eV ranges respectively.  The reach at the shot-noise limit is formally much stronger than the astrophysical bounds; a careful treatment of the readout-chain noise floor (as begun in~\cite{heterodyne-photon}) is required for realistic projections.  Feasibility studies of such upgrades are warranted for both detectors, as are further studies of the LIGO and Cosmic Explorer arm-cavity based approach of~\cite{Nagano-2019,Nagano-2021}.

The birefringence signal in each of these configurations carries characteristic annual and sidereal modulation signatures from the Galactic DM wind, which serve as clean discrimination tools against terrestrial systematics.  Local DM substructure with anisotropic velocity dispersion can shift the peak sensitivity mass and enhance the reach in specific mass windows.  For any candidate detection, cross-correlation of the modulation pattern across multiple detectors (both terrestrial and space-based) provides a definitive test of the DM origin.

\vskip 0.2in
\noindent \emph{Note added.}  After submission of the original version of this manuscript, I became aware---through the response of colleagues and through the report of the referee---of a substantial body of prior literature on interferometric detection of axion-induced photon birefringence. The present manuscript has been revised throughout to acknowledge this prior work, and to highlight how the present work both complements and builds on previous analyses.  I am grateful to several colleagues who brought their work to my attention, and to the referee for a detailed and constructive report that resulted in this revised and corrected version.

\vskip 0.2in
I acknowledge numerous discussions and suggestions during the formative stages of this work with Jayden Newstead and Subir Sabharwal, who also commented on the draft manuscript and confirmed several of the quoted results.  I also acknowledge information from Anthropic CLAUDE Opus 4.7 regarding details of practical heterodyne photodetector technology, and numerical verification of estimates made of detector sensitivities in both the audio and heterodyne regimes.


\begin{thebibliography}{99}

\bibitem{pq}
Peccei, R. D. and Quinn, H. R., Phys. Rev. Lett. 38, 1440  (1977), S. Weinberg, Phys. Rev. Lett. 40, 223 (1978), F. Wilczek, Phys. Rev. Lett. 40, 279 (1978).

\bibitem{dm}
Abbott, L. and Sikivie, P., Phys. Lett. B120, 133 (1983), Preskill, J., Wise, M.B., Wilczek, F., Phys. Lett B120, 127 (1983), Dine, M., Fischler, W.. Phys.Lett. B120 (1983) 137.

\bibitem{sikivielmk} 
P. Sikivie Phys. Rev. Lett. 52, 695 (1984),
L. M. Krauss, J. Moody, F. Wilczek, D. E. Morris,  Phys. Rev. Lett 55(17), 1797-1800 (1985).

\bibitem{DM-Radio} DM Radio Collaboration,
  Phys.\ Rev.\ D \textbf{106}, 103008 (2022).

\bibitem{ABRACADABRA} ABRACADABRA-10cm Collaboration,
Phys.\ Rev.\ Lett.\ \textbf{122},
 121802 (2019).

\bibitem{MADMAX} MADMAX Collaboration,
 Eur.\ Phys.\ J.\ C \textbf{79}, 186 (2019).

\bibitem{Krauss-2019} L.~M.~Krauss, arXiv:1905.10014 [hep-ph] (2019).

\bibitem{renau}
Espriu, D and Renau, A., Phys. Rev. D85, 025010 (2012), Renau, A., arXiv  e print 1512.03311 (2015) .

\bibitem{LiuSmoot} T.~Liu, G.~Smoot, and Y.~Zhao,
 Phys.\ Rev.\ D \textbf{101}, 063012 (2020).

\bibitem{Parkes-PPA-2024} Parkes Pulsar Timing Array Collaboration,
 arXiv:2412.02229 (2024).

\bibitem{EPTA-PPA-2025} European Pulsar Timing Array Collaboration,
 Phys.\ Rev.\ D \textbf{111}, 062005 (2025).

\bibitem{Ivanov} M.~M.~Ivanov \textit{et al.}, JCAP \textbf{02}, 059 (2019).

\bibitem{Fujita-Tazaki} T.~Fujita, R.~Tazaki, and K.~Toma, Phys.\ Rev.\ Lett.\ \textbf{122}, 191101 (2019).

\bibitem{DeRocco-Hook} W.~DeRocco and A.~Hook, Phys.\ Rev.\ D \textbf{98}, 035021 (2018), arXiv:1802.07273 [hep-ph].

\bibitem{Obata-2018} I.~Obata, T.~Fujita, and Y.~Michimura, Phys.\ Rev.\ Lett.\ \textbf{121}, 161301 (2018), arXiv:1805.11753 [astro-ph.CO].

\bibitem{Liu-Elwood} H.~Liu, B.~D.~Elwood, M.~Evans, and J.~Thaler, Phys.\ Rev.\ D \textbf{100}, 023548 (2019), arXiv:1809.01656 [hep-ph].

\bibitem{Martynov-Miao} D.~Martynov and H.~Miao, Phys.\ Rev.\ D \textbf{101}, 095034 (2020), arXiv:1911.00429 [physics.ins-det].

\bibitem{DANCE} Y.~Oshima \textit{et al.}, Phys.\ Rev.\ D \textbf{108}, 072005 (2023), arXiv:2303.03594 [hep-ex].

\bibitem{ADBC} S.~Pandey, E.~D.~Hall, and M.~Evans, Phys.\ Rev.\ Lett.\ \textbf{133}, 111003 (2024), arXiv:2404.12517 [hep-ex].

\bibitem{LIDA} J.~Heinze \textit{et al.}, Phys.\ Rev.\ Lett.\ \textbf{132}, 191002 (2024), arXiv:2307.01365 [hep-ex].

\bibitem{Nagano-2019} K.~Nagano, T.~Fujita, Y.~Michimura, and I.~Obata, Phys.\ Rev.\ Lett.\ \textbf{123}, 111301 (2019), arXiv:1903.02017 [hep-ph].

\bibitem{Nagano-2021} K.~Nagano, H.~Nakatsuka, S.~Morisaki, T.~Fujita, Y.~Michimura, and I.~Obata, Phys.\ Rev.\ D \textbf{104}, 062008 (2021), arXiv:1912.09123 [hep-ph].

\bibitem{DarkGEO} J.~Heinze \textit{et al.}, New J.\ Phys.\ \textbf{26}, 055002 (2024), arXiv:2401.11907 [hep-ex].

\bibitem{Gue-2025} J.~Gu\'e, A.~Hees, and P.~Wolf, Class.\ Quantum Grav.\ \textbf{42}, 055015 (2025), arXiv:2410.17763 [hep-ph].

\bibitem{Yao-2025} Y.-H.~Yao, T.~Jiang, and Y.~Tang, Phys.\ Rev.\ D \textbf{111}, 055031 (2025), arXiv:2410.22072 [hep-ph].

\bibitem{heterodyne-photon} H.~Yu, O.~Kwon, D.~K.~Namburi, R.~H.~Hadfield, H.~Grote, and D.~Martynov, Phys.\ Rev.\ D \textbf{109}, 095042 (2024).

\bibitem{LIGO-scalarDM} LIGO Scientific and Virgo Collaborations, Phys.\ Rev.\ Lett.\ \textbf{133}, 101001 (2024), on scalar DM constraints from LIGO.

\bibitem{clusters} C. A. J. OHare, G. Pierobon, and J. Redondo3, Phys. \ Rev. \ Lett. \textbf{133}, 081001 (2024)

\bibitem{structure} J.W. Foster, N. L. Rodd, and B. R. Safdi, Phys. \ Rev. \ D \textbf{97}, 123006 (2018)

\bibitem{Babak-noise} S.~Babak, M.~Hewitson, and A.~Petiteau, arXiv:2108.01167 [astro-ph.IM] (2021).

\bibitem{CAST} CAST Collaboration,
  Nature Phys.\ \textbf{13}, 584 (2017).

\bibitem{MWDPolarisation} C.~Dessert, D.~Dunsky, and B.~R.~Safdi, Phys.\ Rev.\ D \textbf{105}, 103034 (2022).

\bibitem{Chandra-NGC1275} C.~S.~Reynolds \textit{et al.}, Astrophys.\ J.\ \textbf{890}, 59 (2020).

\bibitem{AxionLimits} C.~O'Hare, \texttt{cajohare/AxionLimits}, Zenodo, doi:10.5281/zenodo.3932430, \url{https://cajohare.github.io/AxionLimits/} (2020+).

\end{thebibliography}
\end{document}